\documentstyle[aas2pp4]{article}

\slugcomment{}

\lefthead{HEAT Collaboration}
\righthead{Measurements of the Cosmic Ray Positron Fraction From  1 Gev to 50 
GeV}

\begin{document}
\input epsf

\begin{flushright}
UM HE 97-04
\end{flushright}

\vskip 0.4cm

\title{Measurements of the Cosmic-Ray Positron Fraction \\
 From 1 to 50 GeV}

\author{S. W. Barwick,\altaffilmark{1} J.~J.~Beatty,\altaffilmark{2}  
A.~Bhattacharyya,\altaffilmark{3} C.~R.~Bower,\altaffilmark{3} 
C.~J.~Chaput,\altaffilmark{4} S.~Coutu,\altaffilmark{4} 
G.~A.~de~Nolfo,\altaffilmark{2,5} J.~Knapp,\altaffilmark{6} 
D.~M.~Lowder,\altaffilmark{7} 
S.~McKee,\altaffilmark{4} D.~M\"uller,\altaffilmark{8} 
J.~A.~Musser,\altaffilmark{3} S.~L.~Nutter,\altaffilmark{9} 
E.~Schneider,\altaffilmark{1} S.~P.~Swordy,\altaffilmark{8} 
G.~Tarl\'e,\altaffilmark{4} A.~D.~Tomasch,\altaffilmark{4} 
E.~Torbet\altaffilmark{8}}

\altaffiltext{1}{Department of Physics, University of California at Irvine, 
Irvine CA 92717} 
\altaffiltext{2}{Department of Physics and of Astronomy and Astrophysics, 104 
Davey Laboratory, Pennsylvania State University, University Park, PA 16802}
\altaffiltext{3}{Department of Physics, Swain Hall West, Indiana University, 
Bloomington, IN 47405}
\altaffiltext{4}{Department of Physics, Randall Laboratory, 500 E. University, 
University of Michigan, Ann Arbor, MI 48109-1120}
\altaffiltext{5}{Department of Physics, Washington University, St. Louis, MO 
63130}
\altaffiltext{6}{Institut f\"ur Experimentelle Kernphysik, Universit\"at 
Karlsruhe, Postfach 3640, D 76021 Karlsruhe Germany}
\altaffiltext{7}{Physics Department, 355 LeConte Hall, University of California 
at Berkeley, Berkeley, CA 94720}
\altaffiltext{8}{Enrico Fermi Institute and Department of Physics, 933 E. 56th 
St., University of Chicago, Chicago, IL 60637}
\altaffiltext{9}{Department of Physical Sciences, Eastern New Mexico University, 
Portales, NM 88130}

\begin{abstract}
Two measurements of the cosmic-ray positron fraction as a function of energy 
have been made using the High Energy Antimatter Telescope (HEAT) balloon-borne 
instrument.  The first flight took place from Ft. Sumner, New Mexico in 1994, 
and yielded results above the geomagnetic cutoff energy of 4.5 GeV.  The second 
flight from Lynn Lake, Manitoba in 1995 permitted measurements over a larger 
energy interval, from 1 GeV to 50 GeV.   In this letter we present results on 
the positron fraction based on data from the Lynn Lake flight, and compare these 
with the previously published results from the Ft. Sumner flight.  The results 
confirm that the positron fraction does not increase with energy above $\approx$ 
10 GeV, although a small excess above purely secondary production cannot be 
ruled out. At low energies the positron fraction is slightly larger than that 
reported from measurements made in the 1960's. This effect could possibly be a 
consequence of charge dependence in the level of solar modulation.

\end{abstract}

\keywords{cosmic rays, elementary particles, instrumentation:detectors, 
ISM: abun\-dan\-ces}

\section{Introduction}
Cosmic-ray electrons and positrons interact with the interstellar medium 
exclusively through electromagnetic processes, such as synchrotron radiation and 
inverse Compton scattering, which do not significantly affect the nucleonic 
cosmic-ray components.  For this reason, electrons are a unique probe of 
cosmic-ray confinement and source distribution in the galaxy. The observed 
$e^\pm$  flux is dominated by negative electrons from primary acceleration 
sites.  However, about 10\% of the total flux are secondary particles, resulting 
from hadronic interactions between the nuclear cosmic rays and nuclei in the 
interstellar medium.  These interactions produce electrons and positrons in 
roughly equal numbers.  A number of observations 
(\cite{agrinier,buffington,muller,golden1,golden2}) indicate that the positron 
fraction, $e^+/(e^-+e^+)$,  increases with energy at energies above 10 GeV. Such 
an increase would require either the appearance of a new source of positrons, or 
a depletion of primary electrons.  Confirming either of these possibilities 
would have a profound impact on our understanding of cosmic-ray sources.  This 
motivated the construction of the HEAT (High Energy Antimatter Telescope) 
instrument, which was designed to determine the positron fraction over a wide 
energy range, and with improved statistical and systematic accuracy.  A first 
balloon flight (HEAT-94) was launched in 1994 from Ft. Sumner, New Mexico, at a 
geomagnetic cutoff rigidity of roughly 4.5 GV.  The results from this flight 
(\cite{heat1}) indicate that the positron fraction does not increase at high 
energies, and that positrons may well be of entirely secondary origin.  A 
subsequent publication (\cite{golden3}) by another group reports a positron 
fraction which is statistically consistent with both the HEAT result and the 
previous measurements. 

In this letter we report the positron fraction measured during the second HEAT 
flight (HEAT-95).  This flight was launched in 1995 from a location with a low 
cutoff rigidity (Lynn Lake, Manitoba), in order to permit measurements over a 
larger energy range of 1 GeV up to about 50 GeV.  We present the results from 
this flight and compare this measurement with the previously reported data from 
HEAT-94. The two data sets have been  analyzed in the same fashion, to avoid 
potential systematic differences in the final result.

\section{Detector Description}
The HEAT detector is described in detail elsewhere (\cite{nim}). It has a 
geometrical acceptance of 473 cm$^2$-sr, and consists of a magnetic spectrometer 
combined with a transition radiation detector (TRD), an electromagnetic 
calorimeter (EC), and time-of-flight (ToF) scintillators.  

 The magnetic spectrometer provides particle tracking through an array of drift 
tubes (DT) in the magnetic field generated by a two-coil, warm-bore 
superconducting magnet.  The DT array consists of 19 tracking layers in the 
bending plane, and 8 layers in the non-bending plane, each providing a 
single-point tracking resolution of $\approx 70 \mu \rm m$. The performance of 
the magnetic spectrometer as a whole can be characterized by the maximum 
detectable rigidity (MDR), which is the rigidity at which the momentum of the 
particle is equal to the error in the momentum measurement. The mean MDR 
achieved in both flights is 170 GV for electrons.          
	
While the magnetic spectrometer determines the momentum and the sign of the 
particle charge, additional measurements are needed to find the magnitude of the 
particle charge, the direction of particle traversal through the instrument, the 
energy of electrons and positrons, and to reject hadronic background.  The TRD 
provides electron identification and hadronic background rejection through the 
detection of transition x-rays, which can only be produced by particles 
possessing a large Lorentz factor ($\gamma > 10^3$).  The TRD consists of 6 
layers of proportional chambers and associated radiators.  Using a 
maximum-likelihood technique we determine whether the signals induced on the TRD 
chamber cathodes indicate the presence of TR, or are consistent with ionization 
energy loss alone.  In addition, the anode wires of each TRD chamber are read 
out in 25 ns time slices, in order to detect ionization clusters typical for 
x-ray signals.  The time-slice data are analyzed with a neural-net technique and 
provide additional electron discrimination. The combination of maximum 
likelihood and time-slice analysis achieves a hadron rejection factor of $\geq$ 
100 at an electron efficiency of 90\%. 
	 
The EC consists of 10 layers of Lead and plastic scintillators.  The 10 EC 
signals recorded for an event are used to measure the primary energy of the 
particle,  to determine the degree to which the longitudinal shower development 
matches that expected for an electron, and the depth at which the shower starts. 
 These quantities  are obtained from a covariance analysis, using accelerator 
calibrations of the EC and GEANT/FLUKA-based (\cite{geant,fluka})  
simulations. Over most of the energy range of interest the energy resolution of 
the EC is 10\%.

The event trigger included the requirement that the energy deposited in the last 
7 layers of the EC  exceed that expected for a 0.5 GeV electron during the first 
flight, and a 1 GeV electron during the second flight.  Although this 
requirement reduces the observed numbers of electrons and positrons near the 
threshold energy, it has no effect on the measured positron fraction, since the 
response of the EC to electrons and positrons of the same energy is identical. 
The hadron rejection factor for the trigger and EC is 200 at an electron 
efficiency of 90\%.  An additional energy-dependent hadron rejection factor of 
1.5 to 10 is provided by a comparison between the energy measured in the 
calorimeter with the momentum measured by the spectrometer. The hadron 
rejection factors and electron efficiencies of the EC and TRD are determined 
from the flight data by using the particle identification obtained from one of 
the detectors to define clean samples of negative electrons and protons, and 
applying the electron selection criteria of the other detector to these 
samples.

The ToF system consists of a layer of four plastic scintillators at the top of 
the instrument, and the first 3 scintillator layers of the EC. The top 
scintillators also provide a measurement of the magnitude of the particle charge 
to distinguish singly-charged particles from heavier nuclei. The time-of-flight 
measurement is used to eliminate upward-going (albedo) particles, which mimic 
antiparticles in the spectrometer.  The rejection power of the ToF system is 
sufficient to reduce the  Helium and albedo background to negligible levels.   
           
\section{Flight Summaries}

The first HEAT flight, HEAT-94, took place from Ft. Sumner, NM, on 1994 May 3.  
The total time at float altitude was 29.5 hours, with a mean atmospheric 
overburden of 5.7 g/cm$^2$.  During the flight, the geomagnetic cutoff rigidity 
at the detector varied between  4 GV and 4.5 GV.  The second HEAT flight, 
HEAT-95, took place from Lynn Lake, Manitoba, on 1995 August 23.  During this 
flight, the total time at float altitude was 26 hours.  The mean atmospheric 
overburden was 4.8 g/cm$^2$, and the geomagnetic cutoff rigidity at the detector 
was well below 1 GV.  The detector configuration was essentially the same for 
the two flights.  In the HEAT-95 flight, one of the TRD chambers was inoperative 
due to a high-voltage system failure.  The resulting loss of hadron rejection 
power is small, and has been compensated for in the HEAT-95 analysis by slightly 
tighter electron selection criteria and correspondingly lower electron 
efficiency. All systems achieved comparable levels of performance in the two 
flights.

\section{Data analysis}

Both data sets are subjected to essentially identical selection criteria to 
obtain a final sample of $e^\pm$.  A complete list of these selections is shown 
in Table 1. The first set selects for singly-charged, downward-going particles 
which have a well-determined momentum, and a velocity measurement consistent 
with a $\beta=1$ particle.  The second set of criteria selects for $e^\pm$. The 
TRD maximum-likelihood and time-slice analyses, in addition to the shower-shape 
analysis obtained from the EC, result in a clean sample of $e^\pm$. An energy 
and momentum selection appropriate to the geomagnetic rigidity cutoff of each 
flight is then made.  Finally, we require that the energy, \it E \rm, measured 
in the EC be consistent with the momentum, \it p \rm, determined with the 
magnetic spectrometer.  We evaluate the measured distributions of the ratio \it 
E/p \rm in order to determine the  residual background in the data sets. The \it 
E/p \rm ratio should be unity for $e^\pm$, (subject to instrumental resolution), 
but will normally have a value less than unity for hadrons.   The \it E/p \rm 
distributions obtained for the two flights are shown in Figure 1. The  \it E/p 
\rm interval used in the final data sets is shown as the cross-hatched region in 
this figure. The events falling outside this interval are primarily 
interacting hadrons which have survived the EC and TRD electron selections, 
along with electrons and positrons falling in the low-side tail of their \it E/p 
\rm distribution.  Applying the \it E/p \rm selection symmetrically 
ensures that the positron fraction is not biased by this selection. The 
level of background in the positron sample is estimated by determining the shape 
of the \it E/p \rm distribution for interacting hadrons as well as that for 
electrons, and by fitting the measured \it E/p \rm distribution for positron 
candidates to a superposition of these hadron and electron distributions for 
each \it E/p \rm interval.  This background represents  $\approx$ 1\% of the 
positron counts at low energies, increasing to almost 10\% at high energies.

\placetable{tbl-1}

\begin{figure}
\vspace{-0.5cm}
\hbox to \textwidth{\hss
\epsfxsize=0.4\textwidth
\epsfbox[175 265 420 570]{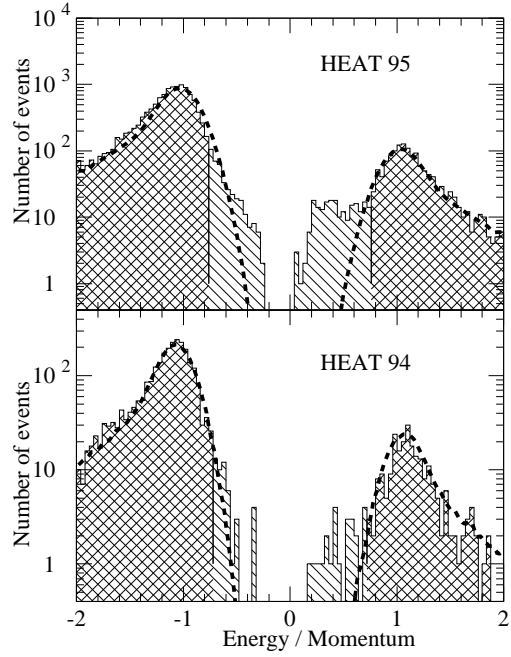}
\hss}
\caption{\it E/p \rm distributions for $e^\pm$ selection.  The 
dashed curve is a Monte Carlo calculation of the expected distributions.  The 
cross-hatched region indicates the \it E/p \rm interval used in the final data 
sample. Note that the vertical scale is logarithmic.\label{fig1} }
\end{figure}

\placefigure{fig1}

Table 2 shows the corrected positron and electron counts and resulting positron 
fractions obtained from this analysis for the two data sets, binned according to 
the energy of the particle. The particle energy has been corrected for radiative 
losses to the top of the atmosphere.  The corrected electron and positron counts 
shown in Table 2 are obtained by subtracting the hadronic background and the 
secondary positrons and electrons produced in the atmosphere from the raw 
counts. The atmopheric contribution is determined by a Monte Carlo simulation of 
hadronic interactions of cosmic rays in the atmosphere.  Over most of the energy 
range of interest the flux of positrons and electrons produced in the atmophere 
is found to be $\approx$ 3\% of the total $e^\pm$ flux.  An empirical estimate 
of the atmospheric contribution is also obtained from the flight data by 
comparing  the positron fraction measured at depths less than 4 g/cm$^2$ to that 
measured at depths greater than 6 g/cm$^2$ as a function of energy.  The 
atmospheric corrections to the positron fraction determined from the Monte Carlo 
simulation and from direct measurements agree within the statistical error of 
the direct measurement. The systematic error in the positron fraction resulting 
from uncertainties in the atmospheric background correction is estimated to be 
1\% for energies well above the geomagnetic cutoff.

\placetable{tbl-2}

\section{Results}

The positron fraction as a function of energy is shown in Figure 2 for the two 
data sets, along with results from previous measurements by other groups, the 
predictions for purely secondary positron production,  (\cite{protheroe}), and 
the predicted positron fraction based on recent work (\cite{clem}) which 
investigated the possible effect of charge sign-dependent modulation.  The 
HEAT-94 data points shown in Figure 2 are essentially those previously published 
(\cite{heat1}).  Minor changes, resulting from  a refinement of the atmospheric 
secondary correction, include the elimination of events with energies between 
4.5 and 5 GeV, for which the secondary correction is more uncertain due to the 
proximity of the geomagnetic cutoff. The HEAT-95 measurement reinforces the 
conclusion that the positron fraction does not increase in the 10-50 GeV energy 
range: the results of the two measurements are consistent with each other, and 
with a general decrease of the positron fraction with energy.  Both data sets do 
indicate an overabundance of positrons compared with the prediction of Protheroe 
at all energies, but this disagreement may not be taken too seriously, as the 
model itself has inherent uncertainties. For example, the predicted positron 
fraction scales with the  ratio of the absolute proton to electron flux, and the 
uncertainties in these quantities are directly reflected in the positron 
fraction. The combined data set suggests the presence of a feature in the 
positron fraction in the energy range from 7 to 20 GeV.  Positron production 
mechanisms have been suggested  (e.g. \cite{kamionkowski}) which would lead to 
an excess of positrons in this energy region, but the uncertainties in our data 
do not permit a definite conclusion, and further measurements are required to 
confirm that this feature exists.  

\begin{figure}
\vspace{-2.5cm}
\hbox to \textwidth{\hss
\epsfxsize=0.9\textwidth
\epsfbox{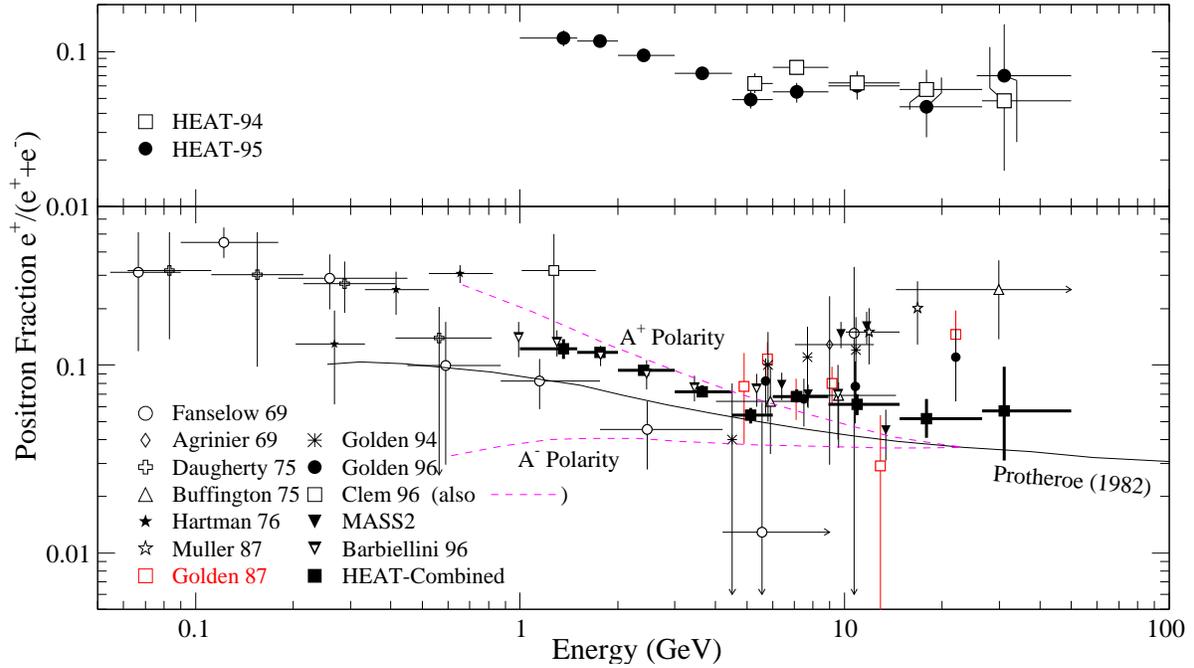}
\hss}
\caption{The positron fraction as a function of energy obtained 
from the two data sets. The error bars applied to the HEAT data points represent 
statistical errors only. \label{fig2}}
\end{figure}

\placefigure{fig2}

At energies below a few GeV, our measured posi\-tron fraction is significantly 
higher than that reported in 1969 by Fanselow et al., while it is in excellent 
agreement with a recent measurement by an independent group 
(\cite{barbiellini}).  As the earlier measurement was performed at a different 
period in the solar activity cycle, it may be tempting to interpret the results 
as being affected by the charge dependence in the solar modulation. This effect, 
and its impact on the positron fraction measured at Earth, has been investigated 
by several groups (\cite{moraal}, \cite{clem}).  Clem et al. have developed a 
model based on the observed systematic difference in the correlation between the 
electron flux measured in space by the ICE instrument and ground-based neutron 
monitor measurements for the 1980 and 1990 solar polarity epochs. This model 
predicts that the present solar epoch favors the transmission of positive charge 
species, resulting in an enhancement in the measured positron fraction over the 
galactic value.  While our  observations may lend some support to the hypothesis 
of charge sign-dependent solar modulation, a definitive test of this hypothesis 
will have to wait until the onset of the next solar epoch, about the year 2000.
\acknowledgments

We gratefully acknowledge assistance from D. Bo\-na\-se\-ra, E. Drag, D. Ellithorpe, 
M. Gebhard, W. Johnson, D. Kouba, R. Northrup, and J. Robbins.  We also thank 
the NSBF balloon crews that have supported the HEAT balloon flights.  This work 
was supported by NASA grants NAG5-5059, NAG5-5069, NAG5-5070, NAGW-5058, 
NAGW-1995, NAGW-2000 and NAGW-4737, and by financial assistance from our 
universities.

\clearpage

\begin{deluxetable}{crr}
\small
\tablecaption{Selection Criteria \label{tbl-1}}
\tablewidth{0pt}
\tablehead{
\colhead{Selection Description}   & 
\colhead{HEAT-94 Selection Range} &
\colhead{HEAT-95 Selection Range}
} 
\startdata
TRD, DT track match & $\mid$Int$_{DT}$-Int$_{TRD}\mid <$ 25 cm\tablenotemark{a} 
&   \nl
Charge=1 &  0.77$e$ $<$ Z $<$ 1.5$e$ &  0.77$e$ $<$ Z $<$ 1.5$e$ \nl
Velocity=$c$ & 0.5$< \beta <$2.0 & 0.8 $< \beta <$ 2.0 \nl
DT track $\chi^2$ &  $\chi^2 < $ 10.0 &  $\chi^2 < $ 10.0 \nl
\# track points & n$_{fit}$ $>$ 9 & n$_{fit}$ $>$ 8 \nl
DTH rigidity error & MDR/$\mid$R$\mid$ $>$ 4 & MDR/$\mid$R$\mid$ $>$ 4 \nl
\nl
TRD $e^\pm$ M.L. & log(M.L) $>$ 2 & log(M.L) $>$ 2 \nl
\# TRD chambers hit & N$_{TRD}$ =6 &  N$_{TRD}$ =5 \nl
TRD time slice & Neural net output $>$ 0.5 & Neural net output $>$ 0.9 \nl
EC shower shape & $\chi^2_{EC} < 1.8$ & $\chi^2_{EC} < 1.8$ \nl
EC shower start & X$_{start} < 0.8$ r.l. &  X$_{start} < 0.5$ r.l. \nl
Energy, momentum selection & E $>$ 3 GeV, $\mid p \mid >$ 2.5 GeV/c &  E $>$ 1 
GeV, $\mid p \mid >$ 1 GeV/c \nl
$\mid E/p \mid$ & 0.7 $<$ $\mid E/p \mid $ $<$ 3.0 &  0.75 $<$ $\mid E/p \mid$ 
$<$ 3.0 \nl
\enddata
\tablenotetext{a}{Int$_{DT}$ and Int$_{TRD}$ are the intercepts obtained in a 
fit of particle track in the  DT and TRD systems}
\end{deluxetable}

\begin{deluxetable}{ccrrrrrrr}
\small
\tablecaption{$e^\pm$ results \label{tbl-2}}
\tablewidth{0pt}
\tablehead{
\colhead{Energy (GeV)} &
\colhead{E$_{mean}$ (GeV)} &
\colhead{n$_{e^+}^{94}$ } &
\colhead{n$_{e^-}^{94}$ } &
\colhead{n$_{e^+}^{95}$ } &
\colhead{n$_{e^-}^{95}$ } &
\colhead{$f^{94}$  \tablenotemark{a}} &
\colhead{$f^{95}$ }&
\colhead{$f^{Comb.}$}
} 
\startdata
1.0 - 1.5&1.36 & & & 65.9  & 475.3  & & 0.122$\pm 0.016$ & 0.122$\pm 0.016$  \nl
1.5 - 2.0&1.76 & & & 236.3 & 1780.2 & & 0.117$\pm 0.008$ & 0.117$\pm 0.008$  \nl
2.0 - 3.0&2.46 & & & 342.6 & 3300.6 & & 0.094$\pm 0.005 $ & 0.094$\pm 0.005$ \nl
3.0 - 4.5&3.64 & & & 205.5 & 2631.6 & & 0.072$\pm 0.005 $ & 0.072$\pm 0.005$ \nl
4.5 - 6.0&5.14 &48.6 & 730.5 & 62.6 & 1218.6 & 0.062$\pm 0.010$ & 0.049$\pm 
0.006$ & 0.054$\pm 0.006$ \nl
6.0 - 8.9&7.20 & 90.1 & 1049.3 & 48.8 & 846.5 & 0.079$\pm 0.008$ & 0.055$\pm 
0.008$  & 0.068$\pm 0.006 $ \nl
8.9 - 14.8&11.0 & 38.6 & 571.9 & 20.0 & 455.6 & 0.063$\pm 0.010$ & 0.060$\pm 
0.012$  & 0.062$\pm 0.008$ \nl
14.8 - 26.5&18.4 & 13.7 & 227.8 & 6.9 & 148.1 & 0.057$\pm^{0.019}_{0.014}$ & 
0.044$\pm^{0.025}_{0.016}$  & 0.052$\pm 0.013$ \nl
26.5 - 50.0&32.3 & 2.1 & 41.2 & 2.1 & 29.4 & 0.048$\pm^{0.059}_{0.027}$ & 
0.070$\pm^{0.081}_{0.045}$  & 0.057$\pm^{0.042}_{0.027}$ \nl
\enddata
\tablenotetext{a}{$f = e^+/(e^-+e^+)$ }
\end{deluxetable}

\newpage

\onecolumn

\end{document}